# MORPHOLOGY AND KINEMATICS OF $\eta$ CARINAE

W.J. Duschl[1,2], K.-H. Hofmann[1], F. Rigaut[3], and G. Weigelt[1]

1: Max-Planck-Institut für Radioastronomie, Auf dem Hügel 69,
D-53121 Bonn, Germany
2: Institut für Theoretische Astrophysik, Universität Heidelberg,
Im Neuenheimer Feld 561, D-69120 Heidelberg, Germany
3: Canada-France-Hawaii Telescope, P.O.Box 1597,
Kamuela, Hawaii 96743, USA

RESUMEN

We present a high-resolution image of $\eta$ Car. Together with IR and visual observations of the central arcsecond, we use this to discuss the morphological structure of $\eta$ Car on the different length scales. We identify three different structural components: a bipolar outflow, an equatorial disk of streamers, and the speckle objects. We discuss models for the kinematics of the whole complex, and propose observations that could settle the question of the structure of $\eta$ Car.

ABSTRACT

We present a high-resolution image of $\eta$ Car. Together with IR and visual observations of the central arcsecond, we use this to discuss the morphological structure of $\eta$ Car on the different length scales. We identify three different structural components: a bipolar outflow, an equatorial disk of streamers, and the speckle objects. We discuss models for the kinematics of the whole complex, and propose observations that could settle the question of the structure of $\eta$ Car.

## 1. INTRODUCTION

$\eta$ Carinae is certainly one of the most enigmatic objects on the sky (see, e.g., Walborn et al. 1978, Davidson et al. 1986, van Genderen and The 1984, Viotti et al. 1989). Since its big outburst one and a half centuries ago, it has been under continuous detailed observation. Considering this it is surprising how little is known even about its basic properties. In the following, by "$\eta$ Car" we mean both, the central star and the circumstellar environment (Homunculus).

In this paper, we present a high resolution image of $\eta$ Car made in 1985 (Sect. 2). Together with IR and visual observations of the innermost arcsecond around the star itself, we use it as the basis of a discussion of the morphology of $\eta$ Car on different length scales (Sect. 3). We introduce a possible model for its structure and kinematics (Sects. 4 and 5), and propose observations that may help to decide the question for $\eta$ Car's and the Homunculus' structure and kinematics (Sect. 6).

## 2. OBSERVATIONS

We observed $\eta$ Car and the Homunculus with the ESO/MPG 2.2m telescope through an edge filter RG 715. We took 200 exposures of 0.25 sec exposure time each, re-centered and averaged them. The result is shown in figure 1. For our reasoning in the following we will use additional IR images of the central 2" (for details, see Rigaut's paper in this volume), and observations of the speckle objects at 820 and 850 nm (Weigelt & Ebersberger 1986, Hofmann & Weigelt 1988, 1993).

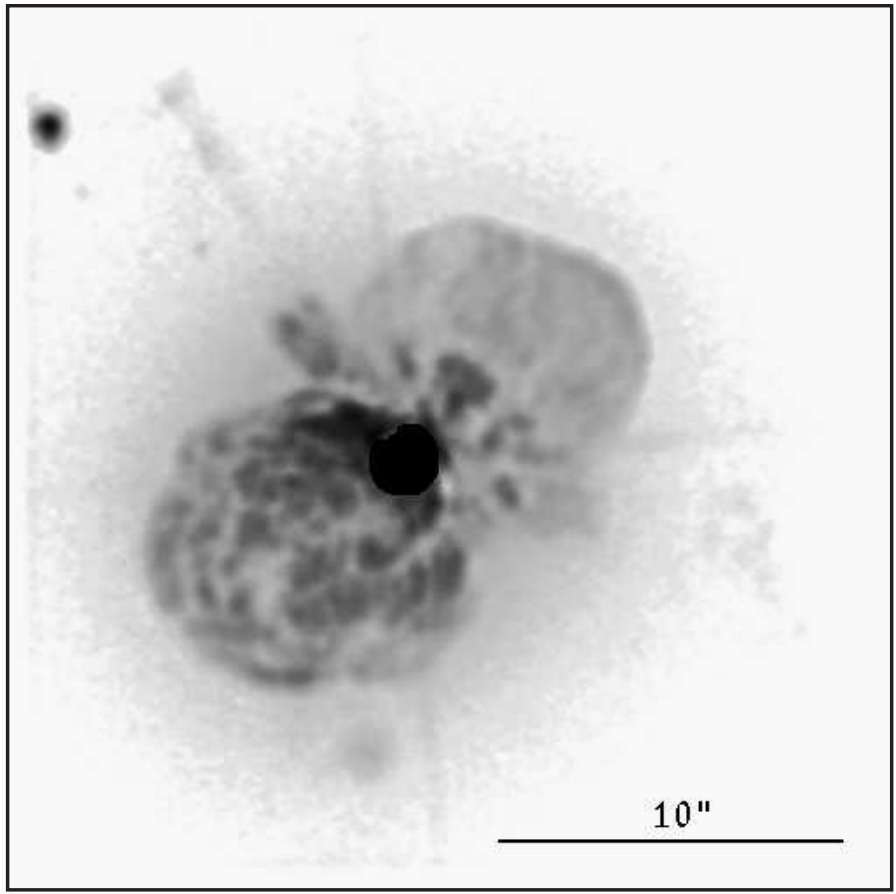

Fig. 1.— Long-exposure image of $\eta$ Carinae obtained by averaging 200 re-centered short exposures. For details see Sect. 2.

## 3. THE MORPHOLOGY OF $\eta$ CARINAE

Combining the information from the above observations, we identify at least three distinct morphological structures of $\eta$ Car, on three different lengths scales (see Fig. 2):

### 3.1. The Homunculus

The famous *Homunculus* has a major axis of 17", with a position angle of 132° (Meaburn et al. 1993a). Although its structure is not yet fully clear, it seems to us that the most likely interpretation is that of a bipolar-like outflow. Given the published body of spectra we cannot yet decide whether this outflow is more like two coni or more like quasi-spherical blobs.

### 3.2. The Equatorial Disk and its Streamers

What used to be adressed as the *Jet* of $\eta$ Car (e.g., Walborn et al. 1978, Hester et al. 1991, Meaburn et al. 1993b) with an orientation perpendicular to the major axis of the Homunculus into the NE direction seems rather to be one of several (at least 8 within the disk's nearer side) radial *Streamers* in an *Equatorial Disk*-like structure between the two parts of the Homunculus outflow. To our knowledge, this structure has not been discussed in the literature yet.

It is important to note that there is coherent structure visible in this equatorial disk of streamers. It seems as if there were different "generations" of substructures in the individual streamers. On the one hand one can identify several bright knots within individual streamers. On the other hand, several knots of different streamers seem to have similar distances to $\eta$ Car itself. We have indicated this in the figure by drawing broken lines for individual streamers and by drawing ellipses connecting blobs of different streamers at the same radial distances from the central object. The ellipses for the different sets of blobs all have the same axis ratio indicating this to be a co-planar structure at an inclination of $\approx 62°$ with respect to the sky plane with the nearer side of the equatorial disk being located to the NW of the star (we use the convention that an inclination of $0°$ corresponds to a structure parallel to the sky plane). The angular radius of this equatorial disk is at least 14". Unfortunately, from the data currently available to us it is not possible to determine whether there is an inner edge to this structure (presumably at a few stellar radii) or whether it extends all the way inwards to the star. We will comment on this later.

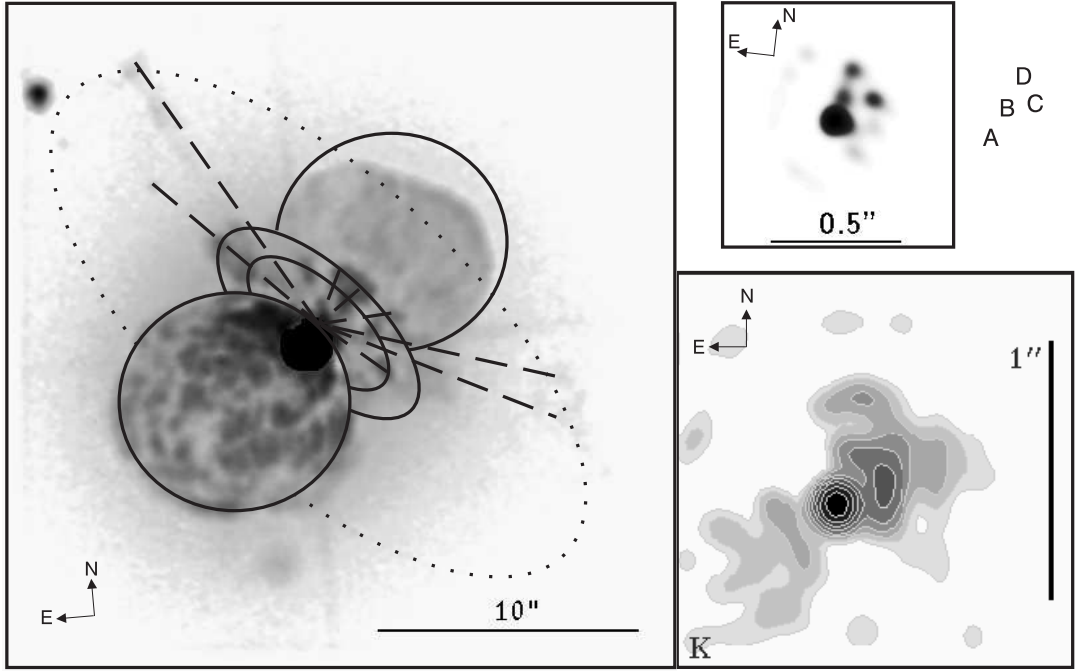

Fig. 2. — Left panel: Same as Fig. 1; broken lines indicating some of the streamers in the equatorial plane; full ellipses indicating the locations of knots in the streamers with similar radial distances from the central object; additionally the two large Homunculus outflows are indicated. Right panels: central arcseconds with the speckle objects; upper panel: 850 nm (from Hofmann & Weigelt 1993); object A is the central object itself, while objects B, C, and D are the speckle objects; lower panel: K-band image (from Rigaut, this volume). The scale in left panel is ten times that of both right panels. The exact directions North are indicated in the individual panels.

In this picture, the often so-called "jet" is only one streamer in the equatorial disk that lies approximately parallel to the sky plane and this way becomes a very prominent feature. At least three other features have almost the same unprojected length as the "jet", but have a weaker contrast to the background. If one accepts this interpretation, it resolves the puzzling problem of a jet perpendicular to a bipolar outflow.

### 3.3. The Speckle Objects

On the smallest observed length scales ($< 1"$), we find the so-called *Speckle Objects* (Weigelt & Ebersberger 1986, Hofmann & Weigelt 1988, 1993) consisting of at least three gas blobs clearly separated from the star. The ratio of the brightnesses and the possibly slightly extended structure of the central object itself makes one wonder whether one sees the star directly or whether it is hidden behind other material and is visible to us only indirectly as scattered or reflected radiation. We will comment on this in the next section when discussing the possible kinematics.

## 4. THE KINEMATICS OF $\eta$ CARINAE

From measurements of the radial and tangential velocities, it seems to be clear that at least the two lobes of the Homunculus as well as the often so-called "jet" to the NE originated in the big outburst of the 1840s (see, e.g., Walborn & Liller 1977, Walborn et al. 1978). If this interpretation is correct, it is an important question how the knots at intermediate distances along the streamers in the equatorial disk relate to the big outburst of the 1840s.

In principle, one may think of two possible interpretations. Either there were *different phases of activity* during which new generations of knots were created and ejected. The advantage of this interpretation is that it offers a most natural explanation of the similar distances at different locations within the equatorial disk. This would require that (1) the blobs of the different generations originated within time scales that are short compared to time intervals between different generations, and that (2) the velocity dispersion of the blobs of one generation is small compared to the ejection velocity. If we further assume that the ejection velocities are the same for all generations of blobs, and that the velocities are constant, then the drawn ellipses would correspond to dates of origin of the blobs of $\approx$ 1840, 1930, and 1950. Interestingly enough, there are no features that can be identified with the "mini-outburst" of 1889 (Walborn & Liller 1977) while Walborn et al. (1978) find structures in the Homunculus that are compatible with an origin around that date. The severe shortcoming of this explanation, on the other hand, is that many of the blobs of different generations seem to align along individual streamers. It seems to us very hard to believe that $\eta$ Car "remembers" the direction from one phase of knot ejection to the next. One would expect such a "memory" only for preferred directions like along the polar axis, but not at all for a equatorial plane where all directions should be equally likely. But, without knowledge about the ejection process itself, even such an unlikely situation cannot be ruled out completely. After all, it is not even clear what creates such streamers in the first place.

The other possibility is that all streamers are due to *one event of ejection* during the big outburst of the 1840s. This gives a very natural explanation for the fact that knots flying away in different directions have a similar distribution of distances from the central object. To achieve such a well structured stacking of the knots, it is necessary that not all the material was ejected at the same velocity but rather had a distribution with certain preferred velocities. While this explanation circumvents the severe problems of the one just discussed, it remains to be explained why the outburst created such a discrete spectrum for the velocities of the ejected material. One possibility is that $\eta$ Car underwent several explosive events within one big outburst (that may even be separated by up to several months) with different velocities of the ejecta from the different explosive events, but similar velocities within individual explosive events. If all the material in the streamers was ejected during a period of time that is short compared to the time since then, then, at any time, we have a linear radial dependence of the velocities. The drawn ellipses of knots at similar distances then correspond to velocities of $\approx$ 1400 km sec$^{-1}$ (Meaburn et al. 1993), 630 km sec$^{-1}$, and 400 km sec$^{-1}$, respectively.

Finally, a combination of the two proposals, or complications in either model (e.g., asymmetries within the equatorial plane, etc.) cannot be ruled out at all. For a recent model for the outbursts in Luminous Blue Variables, see Stother & Chin (1993).

## 5. DISCUSSION

In the next section, we will discuss how the picture described above can be tested and verified (or falsified). In the present Section, we assume this model to apply to $\eta$ Car, and we will discuss in how far it leads to a consistent picture on the different length scales. While the large and medium scale structures (Homunculus and the different sets of knots in the streamers) seem to fit together naturally as polar outflow with the streamers in a plane that is oriented perpendicularly to the major axis of the outflow, the small scale structures need some further comments.

Given the visual and the IR observations on length scales of a few arcseconds and smaller, there are several possibilities. At 850 nm, the three gas blobs are only about 12 times fainter than the star itself (Hofmann & Weigelt 1988). A natural explanation of this situation could be that what we see as the star is not its full light, perhaps not even the star proper. It could very easily be that the star itself is hidden from us by some circumstellar material that is abundantly present in its immediate vicinity. What we see then would be only a image of the star reflected off some of the circumstellar material. The gas blobs, on the other hand, would be fully lit by the star. This could explain the brightness ratios as well as the slightly elongated image of the star in the speckle observations. Moreover, the direction of the elongation is in good with that of the major axis of the equatorial disk (see Fig. 2).

Rigaut (this volume) finds a spectral component corresponding to a temperature of $\approx$ 1000 K and a luminosity of $\approx 10^4$ $L_\odot$. Together with the structure of, e.g., the image in the K band, one could speculate as to whether the obscuring material is an acretion disk around $\eta$ Car. An accretion disk around a 100 $M_\odot$ ($M_*$) star with an inner radius of 30 $R_\odot$ ($s_{\mathrm{inner}}$) and an outer radius $\gg$ 1 AU and $\ll$ 100 AU ($s_{\mathrm{outer}}$), and with a radial mass flow rate inwards of $10^{-4}$ $M_\odot$/year ($\dot{M}$), i.e., reasonable numbers for a system like $\eta$ Car, leads to a luminosity $L_{\mathrm{disk}}$ and an average temperature ($\bar{T}_{\mathrm{disk}}$) close to the observed values:

$$L_{\mathrm{disk}} \approx \frac{GM_*\dot{M}}{s_{\mathrm{inner}}} \approx 4 \cdot 10^{37} \mathrm{erg\ sec}^{-1} \approx 10^4 L_\odot$$

and

$$L_{\mathrm{disk}} \approx \pi s_{\mathrm{outer}}^2 \cdot \sigma \bar{T}_{\mathrm{disk}}^4 \quad \Rightarrow \quad s_{\mathrm{outer}} \approx 30\ \mathrm{AU} \hat{=} 10\ \mathrm{marcsec}$$

In such a picture, one could understand the blobs as being either material that is just streaming out into the bipolar outflow perpendicular to the accretion disk or as material that lies and moves almost in the equatorial plane. In both cases the direct radiation from the star is hidden from us by the disk. The disk may be oriented in the same direction as the bigger equatorial disk with the streamers (but note: the large 14" equatorial disk is not an accretion disk!) and radiation would reach us only as scattered light. Another possibility is that the elongated image is due to reflection off the disk's inner boundary.

To clarify the situation, it is tempting to investigate whether there is some relation between the speckle gas blobs and the streamers on the larger scale.

For the streamers and its knots we discussed the two possibilities of (1) a nearly co-eval origin with a certain velocity distribution, or of (2) the same velocity ($\approx$ 1400 km sec$^{-1}$) for all knots allowing for different epochs of ejection. The latter assumption leads to a doubling time scale of the tangential distances between the speckle blobs B, C, and D and the star (A) of the order of only very few years. This is clearly ruled out by observations (see Weigelt et al., this volume). On the other hand, assuming that all knots are co-eval, those at 0.1...0.2" from the star should have a velocity around a few tens of km sec$^{-1}$. It is interesting to note that these velocities are in agreement with observations of the blobs' proper motions. But one has also to note that this by itself is only an indication for but not the proof of a connection between the speckle blobs and the streamers.

Further evidence may be obtained from the position angles of the speckle knots and the streamers. Knots B and D have position angles very similar to the two streamers pointing towards NW almost along the Homunculus' axis. For knot C such a correspondence is not obvious.

## 6. FURTHER OBSERVATIONS

To verify or falsify the above introduced model for $\eta$ Car clearly further observations are required. The most promising one seems to us to be the determination of a high resolution (both, in positional and in velocity resolution) 2D radial velocity field of the Homunculus and the partially overlayed equatorial disk. If the – to us unlikely – possibility is correct that the streamers all have about the same velocity, such a high resolution spectroscopy should show a prominent and easily detectable high radial velocity feature at the place of the equatorial disk. On the other hand, if the knots are co-eval we have a much more unfortunate situation as then the expansion velocities of the Homunculus outflow and of the knots projected onto it may be very similar. Then, the different gradients of radial velocities in the outflow and in the equator region contain the required information.

## 7. CONCLUSIONS

Based on high resolution visual and IR observations of the Homunculus and the innermost region around the star itself, we have discussed several possible models. While we wish to emphasize that the currently available data do not yet allow a clear decision between different possibilities, in our opinion the most likely configuration is that of a bipolar-like outflow along polar axis with an equatorial disk with several (at least eight) expanding streamers in a plane perpendicular to the outflow. It is also not clear how the speckle objects relate to these large scale structures. But there are some indications pointing towards the speckle blobs being material that either is just about to stream away from the star into the outflow or, more likely, connects to the streamers in the equatorial disk. In the last section, we discussed how 2D spectroscopy will allow us to gain further constraints and to, hopefully, identify the 3D configuration of $\eta$ Car, the Speckle Objects, and the Homunculus.